\documentclass[%
 reprint,
 onecolumn,
 amsmath,amssymb,
 pra,
]{revtex4-2}

\usepackage{graphicx}
\usepackage{dcolumn}
\usepackage{bm}

\newcommand{\ket}[1]{|#1\rangle}
\newcommand{\bra}[1]{\langle #1|}

\graphicspath{{./pics/}}

\usepackage[figurename=FIG., justification=RaggedRight]{caption}
\usepackage{subcaption} 
\usepackage{physics}

\usepackage[titletoc]{appendix}

\begin{document}

\title{Explicit attacks on BB84 with distinguishable photons}

\author{D. Babukhin$^{1,2}$}
\author{D. Kronberg$^{2}$}
\author{D. Sych$^{2,3}$}

\affiliation{$^1$QRate LLC, Novaya av. 100, Moscow 121353, Russia}
\affiliation{$^2$Department of Mathematical Methods for Quantum Technologies, Steklov Mathematical Institute of Russian
Academy of Sciences, Gubkina str. 8, Moscow 119991, Russia}
\affiliation{$^3$P.N. Lebedev Physical Institute, Russian Academy of Sciences, 53 Leninskiy Prospekt, Moscow 119991, Russia}

\begin{abstract}

Distinguishability of photons in non-operational degrees of freedom compromises unconditional security of quantum key distribution since an eavesdropper can improve attack strategies by exploiting this distinguishability. However, the optimal eavesdropping strategies in the presence of light source side channels are not known. Here we provide several explicit attack strategies on the BB84 protocol with partially distinguishable photons. In particular, we consider the phase-covariant cloning attack, which is optimal in the absence of side channels, and show that there are even better strategies in the presence of side channels. The improved strategies exploit a measurement of the side channel state followed by an attack on the signal photon. Our results explicitly demonstrate reduction of the critical error rate, and set an upper bound on the practical secret key rate.

\end{abstract}

\maketitle

%
%

\section{Introduction}

Quantum key distribution (QKD) is a method to share a secret key between two legitimate parties (Alice and Bob) in the presence of an eavesdropper (Eve) \cite{gisin2002, Pirandola2020}. This possibility is based on the no-cloning theorem, which prohibits information gain from a carrier photon without disturbing its quantum state \cite{Wooters1982, Dieks1982}. In theory, QKD is unconditionally secure \cite{LoChau1999, Shor2000}. In practice, however, there are numerous deviations of practical QKD from theory \cite{Scarani2014} which leads to loopholes and side channels \cite{Jain2016} and, consequently, to quantum hacking \cite{Makarov2017, Huang2018}. The practical security of QKD is an active research topic \cite{Diamanti2016, Xu2020}.  

Loopholes on Bob's (receiver) side can be closed with the use of measurement-device-independent QKD protocol \cite{MDI}, where measurement devices are excluded from the private space of legitimate sides and transferred to a third (untrusted) party. This protocol allows getting rid of all hacking attacks on measurement devices, which made up the majority of already known hacking strategies in practical QKD \cite{Jain2016}. But this protocol does not close loopholes on Alice's (transmitting) side, which require specific characterization and countermeasures (\cite{Lucamarini2015, Tamaki2016, Nauerth2009, Duplinskii2019})

Light source side channels can be described with a non-operational degree of freedom \cite{Lucamarini2015, Pereira2019} from which Eve gains information.
The efficiency of eavesdropping depends on Eve's choice how to measure and process information from the side channel.
In the BB84 protocol without side channels, the critical quantum bit error rate $11\%$ is an information-theoretic result \cite{Shor2000}, which is also achieved with an explicit eavesdropping strategy \cite{Brus2000}.
In the BB84 with distinguishable photons, the lower bound on the possible secret key rate was provided in \cite{Duplinskii2019}. This result shows a pessimistically low secret key generation rate. The explicit attack strategy, which gives this lower bound of the secret key rate, is unknown. Previous research of this issue \cite{Babukhin2020, Babukhin2021, Sych2021} hasn't reached the secret key rate of \cite{Duplinskii2019}, thus opening a gap between explicit attacks and theoretical secrecy. Therefore investigating explicit attack strategies can help close this gap.

In this paper, we provide explicit eavesdropping strategies on the BB84 protocol in the presence of a binary side channel of the light source. 
The strategies consist of a measurement of the side channel state and an attack on the signal photon.
We investigate two main types of measurements: unambiguous state discrimination (USD) and minimum error measurement (ME). For unambiguous state discrimination of the side channel, there are two options: in the case of a conclusive result, Eve obtains full information about the quantum signal, while in the case of an inconclusive result, Eve performs the standard optimal cloner attack. Therefore, we expect reduction of the critical error rate. For the minimum error side channel measurement, a measurement result introduces a bias in the signal states ensemble probabilities. Eve can use this bias as an additional binary information source or, alternatively, she can use it to adjust the eavesdropping strategy on the signal photon with the use of postselection. We use a Hong-Ou-Mandel (HOM) interference visibility \cite{Duplinskii2019} to estimate information leakage through the side channel and calculate how the critical error rate depends on the HOM visibility.

This paper is organized as follows. 
In Sec. II we formulate the BB84 protocol in the presence of a binary source side-channel and discuss possible strategies for information extraction. 
In Sec. III we formulate three eavesdropping strategies on the protocol, and
in Sec. IV we discuss and conclude our results.

\section{BB84 protocol with distinguishable photons}

\subsection{Side channel model}

In the BB84 protocol, Alice and Bob use two bases of quantum states to distribute bits of a secret key: 
X-basis $\{\ket{0}_{x}, \ket{1}_{x}\}$ and Y basis $\{\ket{0}_{y}, \ket{1}_{y}\}$. Alice randomly chooses a secret bit value ($0$ or $1$), and randomly chooses a basis (X or Y) and sends a quantum state into communication channel. This leads to the following ensemble of states
\begin{equation}
    \label{ensembleBB84}
    \biggl{\{} \frac{1}{4}: 
        \ket{0_{x}},\text{  }
       \frac{1}{4}:
       \ket{1_{x}},\text{  }
       \frac{1}{4}:
       \ket{0_{y}},\text{  }
       \frac{1}{4}:
       \ket{1_{y}}
      \biggl{\}}.
\end{equation}
Eavesdropping introduces mixedness in this ensemble, which occurs due to entangling the state of the carrier photon with an auxiliary quantum system. Through measuring her quantum system, Eve obtains outcomes, which correlate with secret bits. At the same time, eavesdropping introduces errors in Bob's measurement results. These errors allow estimating the leakage of information to the eavesdropper. This eavesdropping-errors connection is guaranteed with the no-cloning theorem and is a cornerstone of the quantum key distribution.

The eavesdropper has an informational side channel when the photon source produces photons with physical distinguishability besides the signal degree of freedom.
In this work, we consider a specific binary model of the side channel of the photon source. 
This model accounts for a photon distinguishability of Alice's photon source while providing Eve no information neither about secret bit nor about basis choice.
When Eve measures a state of a side channel, she finds that one couple of quantum states is more probable in a quantum channel. 
At the same time, these states belong to different bases and encode different bits, thus, this is not a side channel of the form ``more 0 than 1''. 
Thus, our model is more interesting from the point of eavesdropping possibilities: even such an uninformative side channel allows Eve to increase eavesdropping efficiency.
The model of side channel is 
\begin{equation}
\label{sch1}
    \biggl{\{}
        \ket{0_{x}}\otimes\ket{0_{\Delta}},\text{  }
        \ket{1_{x}}\otimes\ket{1_{\Delta}}\text{  }
    \biggl{\}}
\end{equation}
for X basis and 
\begin{equation}
\label{sch2}
    \biggl{\{}
        \ket{0_{y}}\otimes\ket{1_{\Delta}},\text{  }
        \ket{1_{y}}\otimes\ket{0_{\Delta}}\text{  }
    \biggl{\}}
\end{equation}
for Y basis, where $\ket{0_{\Delta}}$ and $\ket{1_{\Delta}}$ are two non-orthogonal states $\bra{0_{\Delta}}\ket{1_{\Delta}} = \Delta$ of a side channel degree of freedom. This form of side channel allows for special eavesdropping strategies, which we describe in following sections.

%
%

\subsection{Side channel states measurements strategies}
\label{section3}

Knowledge of the photon distinguishability side channel allows Eve a variety of strategies to attack the protocol. The choice here is how to combine measurements of the signal and side-channel degrees of freedom to gain information about secret bits. In the following, we consider two strategies of interest.

\subsubsection{Unambiguous state discrimination measurement}
\label{subsection3A}

With this strategy, Eve performs unambiguous state discrimination of the side-channel degree of freedom. This measurement provides full knowledge on the side channel state or yields an inconclusive result with no information.
In terms of POVM operators, this measurement is formulated as following: for a set of linearly independent quantum states $\{\ket{\psi_{1}}, ... , \ket{\psi_{n}}\}$ there exist POVM operators $M_{i}, i = 0, ..., n$, $I = \sum_{i=0}^{n}M_{i}$, such that
\begin{equation}
    p_{i} = Tr[\ket{\psi_{i}}\bra{\psi_{i}} M_{i}], \quad Tr[\ket{\psi_{j}}\bra{\psi_{j}} M_{i}] = 0, \text{ if } j \neq i.
\end{equation}
is a probability of conclusive results of measuring $i-th$ state if Eve has a conclusive measurement, she obtains full information about the measured state, and 
\begin{equation}
    p^{inc} = 1 - \sum_{i=1}^{N}p_{i}
\end{equation}
is a probability of an inconclusive measurement with no information about the state.

\subsubsection{Minimum error measurement}
\label{subsection3B}

With this strategy, Eve makes a minimum error measurement of the side-channel degree of freedom, which gives her information of the side channel state with a minimum (but non-zero for non-orthogonal states) error probability. In terms of POVM operators, this measurement is formulated as follows: for a set of quantum states $\{q_{1}: \ket{\psi_{1}}, ... , q_{n}: \ket{\psi_{n}}\}$, where $q_{i}$ is a probability of sending an $i-th$ state towards the measurement device, there exist POVM operators $M_{i}, i = 1, ..., n$, $I = \sum_{i=1}^{n}M_{i}$, such that
\begin{equation}
    p_{i} = Tr[\ket{\psi_{i}}\bra{\psi_{i}} M_{i}]
\end{equation}
is a correct outcome probability when measuring the $i-th$ state and 
\begin{equation}
    p^{error}_{i} = \sum_{\substack{j\neq i \\ j = 1}}^{n} Tr[\ket{\psi_{j}}\bra{\psi_{j}} M_{i}] \neq 0
\end{equation}
is a probability of an error for $i-th$ state measurement.
A condition of minimal error formulates as a constraint of the maximal average probability of correct measurement outcome
\begin{equation}
    P_{opt} = \max_{\{M_{i}\}}\sum_{i=1}^{n}q_{i}p_{i}.
\end{equation}

%
%

\subsection{Soft filtering of an ensemble of quantum states}
\label{section4}

One of the main assumptions in the base of BB84 protocol security is a random choice of the basis and the secret bit, which Alice uses to encode the secret bit into the state of photon and to send it to Bob. Formally this lack of information means equal probabilities of ensemble states (\ref{ensembleBB84}). With source side channels, eavesdropper can obtain information about states from non-operational degrees of freedom. This leads to the equiprobability violation of ensemble states. If states of the ensemble have different probabilities, Eve can use soft filtering \cite{kronberg2020quantum, kronberg2021increasing} to enhance her attack efficiency on the protocol. The soft filtering is an extension of the USD measurement, which in case of success maps 
non-orthogonal quantum states of the ensemble $\{q_{i}: \ket{\psi_{i}}\bra{\psi_{i}} \text{, i = 1, ..., n}\}$ into another ensemble $\{q_{i}\bra{\psi_{i}}Q^{-p}\ket{\psi_{i}}: \frac{Q^{-\frac{p}{2}}\ket{\psi_{i}}\bra{\psi_{i}}Q^{\frac{p}{2}}}{\bra{\psi_{i}}Q^{-p}\ket{\psi_{i}}} \text{, i = 1, ..., n}\}.$, where $Q = \sum_{j}q_{j}\ket{\psi_{j}}\bra{\psi_{j}}$. This quantum channel is defined as follows:
\begin{equation}
    \label{forwardfiltering}
    \Phi(\ket{\psi_{i}}\bra{\psi_{i}}) = F_{succ}\ket{\psi_{i}}\bra{\psi_{i}} F^{\dagger}_{succ} + 
    F_{fail}\ket{\psi_{i}}\bra{\psi_{i}} F^{\dagger}_{fail}
\end{equation}
where
\begin{equation}
    \label{Fsucc}
    F_{succ} = Q^{-\frac{p}{2}},
\end{equation}
\begin{equation}
    F_{fail} = (I - Q^{-p})^{\frac{1}{2}},
\end{equation}
In case of successful filtering, new ensemble consists of states, which are more distinguishable when compared to the states of the initial ensemble. The inverse filtering map is given by 
\begin{equation}
    B_{succ} = Q^{\frac{p}{2}},
\end{equation}
\begin{equation}
    B_{fail} = (I - Q^p)^{\frac{1}{2}},
\end{equation}

Using the described soft filtering, Eve can enhance her eavesdropping efficiency when the photon source has a binary side channel as introduced in (\ref{sch1}) and (\ref{sch2}). 
If Eve makes the minimum error measurement of side-channel degree of freedom, she efficiently introduces bias in quantum states of Alice ensemble. Initial ensemble (\ref{ensembleBB84}) transforms to
\begin{eqnarray}
    \label{bb84ensemble_q}
    \biggl{\{} \frac{q}{2}: 
        \ket{0_{x}},\text{  }
       \frac{1-q}{2}:
       \ket{1_{x}},\text{  }
       \frac{1-q}{2}:
       \ket{0_{y}},\text{  }
       \frac{q}{2}:
       \ket{1_{y}}
      \biggl{\}}
\end{eqnarray}
if Eve measured side channel in a state $\ket{0_{\Delta}}$, or
\begin{eqnarray}
    \label{bb84ensemble_notq}
    \biggl{\{} \frac{1-q}{2}: 
        \ket{0_{x}},\text{  }
       \frac{q}{2}:
       \ket{1_{x}},\text{  }
       \frac{q}{2}:
       \ket{0_{y}},\text{  }
       \frac{1-q}{2}:
       \ket{1_{y}}
      \biggl{\}}
\end{eqnarray}
if Eve measured side channel in a state $\ket{1_{\Delta}}$. Details are in Appendix A section. Then, depending on the side channel measurement, Eve applies soft filtering to make two states out of four possible more distinguishable. This filtering allows her to apply a cloning unitary, tuned on the two most distinguishable states, to attack the protocol. We discuss quantum cloning in the next section.

%
%

\subsection{Quantum cloning}
\label{section5}

\subsubsection{The phase-covariant cloning}
\label{subsection5A}

Among all attack strategies on the state of the signal photon (without taking side channel information into account), the most efficient is a collective unitary attack. The essence of this attack is applying a particular unitary evolution to a system, compound of a signal photon, and an ancillary quantum system in such a way that the resulting entangled quantum state allows for maximal information for Eve at a given error on the Bob side. Eve applies the same unitary operation to all signal states in a sequence of sent bits during Alice and Bob's communication and stores her ancillary system states in a quantum memory. After the bases exchange step, Eve applies a collective measurement to her quantum memory, which gives her a possible maximum of information about distributed secret bits.

A collective unitary attack strategy on the BB84 protocol results in a critical quantum bit error rate of $Q_{c} \approx 11\%$ and can be implemented with the use of a phase-covariant optimal cloning machine \cite{Brus1998}. This cloning machine is unitary of the following form
\begin{eqnarray}
    U\ket{\psi(\phi)}_{B}\ket{0}_{E}\ket{0}_{Anc} = \frac{1}{2}(\ket{0}_{B}\ket{0}_{E}\ket{0}_{Anc} +
    \notag\\
    + \cos\eta\ket{0}_{B}\ket{1}_{E}\ket{1}_{Anc}
    + \sin\eta\ket{1}_{B}\ket{0}_{E}\ket{1}_{Anc} \pm
    \notag\\
    \pm \cos\eta\ket{1}_{B}\ket{0}_{E}\ket{0}_{Anc} \pm 
    \sin\eta\ket{0}_{B}\ket{1}_{E}\ket{0}_{Anc} \pm 
    \notag\\
    \pm \ket{1}_{B}\ket{1}_{E}\ket{1}_{Anc}).
\end{eqnarray}
where $\eta$ is a cloning parameter. When $\eta = 0$, the cloner is identity operator and has no effect on the compound quantum state, and when $\eta = \pi/2$, Eve has Bob's state in her space and Bob's qubit becomes maximally mixed because of entanglement with Eve's ancillary qubit.

The described phase-covariant cloning attack allows for a critical bit error value $Q_{c} \approx 11\%$ for the standard BB84 protocol with two mutually unbiased bases $X$ and $Y$ (or $Z$ and $X$). Photons' distinguishability leads to lower critical error values.

\subsubsection{The two state cloning}
\label{subsection5B}

While optimal phase-covariant cloner allows for the most efficient attack in the absence of passive side channels, it does not use photons' distinguishability, which arises from side-channel states measurements. If Eve measures the side channel state with the minimum error measurement, the probabilities of states in Alice's ensemble change, and Eve knows some states are more probable than others. It is reasonable to use this knowledge to devise an adaptive eavesdropping strategy. 

Such an eavesdropping strategy can be constructed with the use of a two-state optimal cloner machine \cite{Brus2000}. This cloning machine is unitary of the following form:
\begin{equation}
    U\ket{0}_{A}\ket{0}_{E} = 
    a\ket{0}_{A}\ket{0}_{E} + 
    b(\ket{0}_{A}\ket{1}_{E} +
    \ket{1}_{A}\ket{0})_{E} + 
    c\ket{1}_{A}\ket{1}_{E}
\end{equation}
\begin{equation}
    U\ket{1}_{A}\ket{0}_{E} = 
    c\ket{0}_{A}\ket{0}_{E} + 
    b(\ket{0}_{A}\ket{1}_{E} + 
      \ket{1}_{A}\ket{0}_{E}) + 
    a\ket{1}_{A}\ket{1}_{E}
\end{equation}
where 
\begin{equation}
    a = \frac{1}{\cos{2x}}((P + Q \cos{2x})\cos{x} - (P - Q\cos{2x})\sin{x})
\end{equation}
\begin{equation}
    b = \frac{1}{\cos{2x}} P(\cos{x} - \sin{x})\sin{2x}
\end{equation}
\begin{equation}
    c = \frac{1}{\cos{2x}}((P - Q \cos{2x})\cos{x} - (P + Q\cos{2x})\sin{x})
\end{equation}
\begin{equation}
    P = \frac{1}{2}\frac{\sqrt{1 + \sin{2x}}}{\sqrt{1 + \sin^{2}{2x}}}
\end{equation}
\begin{equation}
    Q = \frac{1}{2}\frac{\sqrt{1 - \sin{2x}}}{\sqrt{1 - \sin^{2}{2x}}}
\end{equation}
and $x$ is given by a scalar product $\bra{\psi_{1}}\ket{\psi_{2}} = \sin{2x}$ of two states $\ket{\psi_{1}}$ and $\ket{\psi_{2}}$, which the cloner is tuned on.

%
%

\section{Attack strategies on BB84 the protocol}

Here we provide description of eavesdropping strategies (see Appendix C for calculation of secret key rates).

\subsection{Minimal error measurement of side channel and phase-covariant cloning}

Eve performs a minimal error measurement of the side channel state (Sec. \ref{subsection3B}). She uses information from the side channel as additional information from a classical channel. After measuring the side channel state, she executes a phase-covariant cloning attack (Section \ref{subsection5A}) on the transmitted photon and stores her clone in a quantum memory register. At the end of the communication, Eve obtains basis information for each position in the quantum memory register and makes a collective measurement of the memory register.

\subsection{Minimal error measurement of side channel state, soft filtering and two-state cloning}

Eve applies a minimal error measurement of the side channel state (Sec. \ref{subsection3B}). This measurement gives her partial information about the transmitted state, which transforms states of the BB84 ensemble (\ref{ensembleBB84}). If Eve measured side channel in a state $\ket{0}$, the ensemble (\ref{ensembleBB84}) transforms into (\ref{bb84ensemble_q}), and if she measured side channel in a state $\ket{1}$, the ensemble (\ref{ensembleBB84}) transforms into (\ref{bb84ensemble_notq}). This change of probabilities in the ensemble allows Eve to apply a filtering transformation (\ref{forwardfiltering}), which makes the two states with the highest probabilities more distinguishable while making the other two states less distinguishable. Then, Eve applies a two-state cloning transformation (Section \ref{subsection5B}), which is tuned on more distinguishable pair of states, to a photon in the quantum channel and produces two clones of this photon. Then, she applies a backward filtering transformation to both clones to make them closer to the state sent by Alice. Eve sends one of two identical clones towards Bob and stores another in her quantum memory. At the end of the communication, Eve obtains basis information for each position in the quantum memory register and makes a collective measurement of the memory register.

\subsection{USD measurement of side channel state and phase-covariant cloning}

Eve uses a USD measurement of the side channel state (see Sec. \ref{subsection3A}). 
If the USD measurement is successful, it reliably tells Eve what a side channel state was. If the USD measurement fails, it provides no information about the side channel state. In the case of successful measurement Eve does not attack the signal photon state and waits until the basis exchange between Alice and Bob. The knowledge of the basis reveals to her what quantum state was sent among the two, corresponding to the measured side channel state. In this case, she does not introduce any error in the communication act. In case of measurement failure, Eve executes a phase-covariant cloning attack (Sec. \ref{subsection5A}) on the signal photon and stores her clone in a quantum memory register. At the end of the communication, Eve obtains information about used bases for each position in the quantum memory register and makes a collective measurement of the memory register.

\subsection{Results}

Here we provide calculation results for three eavesdropping strategies.  In Fig.1, we provide critical error rates (error rates for secret key rate R = 0) for different values of the HOM visibility.
\begin{figure}[h!]
	\includegraphics[width=0.99\linewidth]{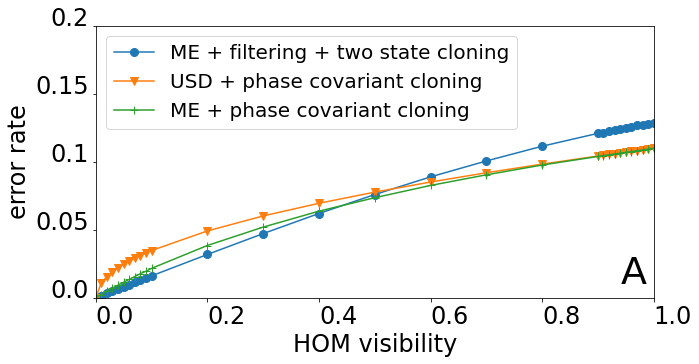}
	\includegraphics[width=0.475\linewidth]{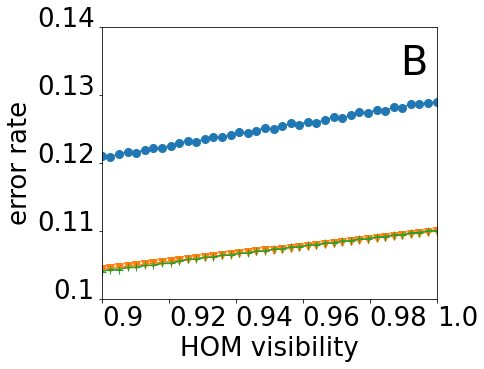}
	\includegraphics[width=0.475\linewidth]{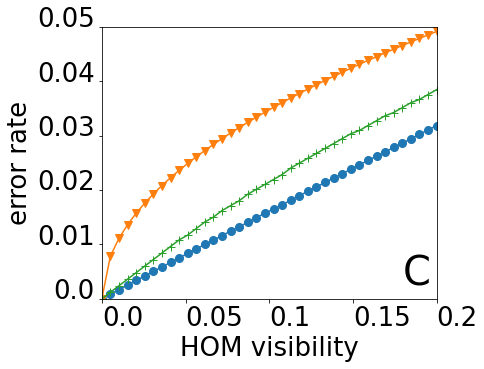}
	\caption{(A) Critical error rates (secret key rate is 0) for three aforementioned eavesdropping strategies. Here Eve attacks side channel state with minimum error measurement (ME) or unambiguous state discrimination (USD). The attack on the side channel state is followed by attack on the signal photon. (B) Critical error rates for high visibility values. (C) Critical error rates for low visibility values. }
\end{figure}
In Fig. 1(A), we see that critical error rates for three strategies have different dependence on the visibility value. The differences in decay rates of critical errors lead to several interceptions of slopes. We see that at $V \approx 0.4$ soft filtering with two-state cloning becomes more efficient than phase-covariant cloning without soft filtering. We also see that the USD measurement on the side channel becomes the least efficient for $V \leq 0.5$. In Fig. 1(B), we see that, for almost perfect single-photon sources, the dependence of critical error rate on visibility is linear for all attack strategies. Contrary, in Fig.1 (C), we see nonlinearity in dependencies of critical error rate on the visibility value (although single-photon sources of this quality are hardly applicable to quantum key distribution). 

\section{Conclusion and outlook}

We compared the efficiencies of three explicit eavesdropping strategies on the BB84 with distinguishable photons. We demonstrated that for light sources with low information leakage (high values of the HOM visibility), it is better to use information from the minimum error measurement of the side channel independently from the attack on the signal photon state to eavesdrop the protocol. In contrast, for light sources with high information leakage (low values of the HOM visibility), it is better to use this information for adaptive eavesdropping with soft filtering of the signal photon state. We found that it is sufficient to use the minimum error measurement of the side channel to reach the best eavesdropping over the whole range of light source HOM visibility values. The USD measurement of the side channel states without blocking some pulses is not better than the other two attacks in our study. 

We also found here that for the BB84 protocol eavesdropping with postselection (soft filtering in our studies) can enhance eavesdropping efficiency compared to eavesdropping without postselection. This enhancement is remarkable since this effect does not take place in the case of BB84 without side channels.

Our results open the following questions for future research. We demonstrated that it is possible to overcome performance of the optimal attack without side channels, but we do not claim the optimality for the proposed attacks. The optimal eavesdropping attack in the presence of side channels as well as the optimal model of side channels remain an open problem. In particular, the dimension of the side channel state space potentially influences the efficiency of eavesdropping; using different photon parameters separately (e.g., spectral and spatial profiles), Eve can construct more complicated sequences of measurements and filtering over the both signal and side channel degrees of freedom, that leads to the lower bound on the secret key rate. 

Next, the critical error rate is not the only possible framework, since there are attacks which compromise the protocol with zero bit error rate. For example, USD of the side channel state with blocking inconclusive results leads to zero error on the Bob's side. To take such strategies into account, we need to include the channel transmittance to the security analysis and connect it to the probability of conclusive results, or, other postselection-based strategies.

\section*{Acknowledgements}

This work was funded by the Ministry of Science and Higher Education of the Russian Federation (grant number 075-15-2020-788).

\renewcommand{\appendixname}{APPENDIX}
\appendix

\section{Ensemble reweighting from minimum error measurement of side channel state}
\label{AppendixA}

Let us consider a process of minimal error measurement of side channel degree of freedom in details. Denote two possible states of side channel degree of freedom as $\{\ket{0_{\Delta}}, \ket{1_{\Delta}}\}$. We model the process of measurement as interaction of a side channel degree of freedom with an ancillary degree of freedom, which states can be distinguished with certainty ($\{\ket{r_{0}}, \ket{r_{1}}\}: \bra{r_{0}}\ket{r_{1}} = 0$). We specify a Stinespring representation of this process as following:
\begin{eqnarray}
    \ket{0_{\Delta}} \longrightarrow \sqrt{P_{succ}}\ket{0_{\Delta}}\ket{r}_{0} + \sqrt{1 -  P_{succ}}\ket{1_{\Delta}}\ket{r}_{1} \\
    \ket{1_{\Delta}} \longrightarrow \sqrt{1 - P_{succ}}\ket{0_{\Delta}}\ket{r}_{0} + \sqrt{P_{succ}}\ket{1_{\Delta}}\ket{r}_{1}    
\end{eqnarray}
where $P_{succ}$ is a probability to distinguish between two states $\ket{0_{\Delta}}$ and $\ket{1_{\Delta}}$. Then, density matrices of two resulting states of the compound system are
\begin{eqnarray}
    \notag
    \rho^{minerr}_{0} = P_{succ}\ket{0_{\Delta}}\bra{0_{\Delta}} \otimes \ket{r_{0}}\bra{r_{0}} +
    \sqrt{P_{succ}(1 - P_{succ})}\ket{0_{\Delta}}\bra{1_{\Delta}} \otimes \ket{r_{0}}\bra{r_{1}} + \\
    \sqrt{P_{succ}(1 - P_{succ})}\ket{1_{\Delta}}\bra{0_{\Delta}} \otimes \ket{r_{1}}\bra{r_{0}} +     
    (1-P_{succ})\ket{1_{\Delta}}\bra{1_{\Delta}} \otimes \ket{r_{1}}\bra{r_{1}} 
\end{eqnarray}
\begin{eqnarray}
    \notag
    \rho^{minerr}_{1} = (1-P_{succ})\ket{0_{\Delta}}\bra{0_{\Delta}} \otimes \ket{r_{0}}\bra{r_{0}} +
    \sqrt{P_{succ}(1 - P_{succ})}\ket{0_{\Delta}}\bra{1_{\Delta}} \otimes \ket{r_{0}}\bra{r_{1}} + \\
    \sqrt{P_{succ}(1 - P_{succ})}\ket{1_{\Delta}}\bra{0_{\Delta}} \otimes \ket{r_{1}}\bra{r_{0}} +     
    P_{succ}\ket{1_{\Delta}}\bra{1_{\Delta}} \otimes \ket{r_{1}}\bra{r_{1}} 
\end{eqnarray}
Now let us look at changes in ensembles of Alice states. Suppose Alice chose an $X$ basis states to send a secret bit to Bob. Eve applies unitary evolution to a side channel state and a measurement device state:
\begin{equation}
    \rho^{X}_{Alice} = \frac{1}{2}\ket{0_{x}}\bra{0_{x}} \otimes \rho_{0}^{minerr} + \frac{1}{2}\ket{1_{x}}\bra{1_{x}} \otimes \rho_{1}^{minerr}
\end{equation}
We can rewrite this state in a more convenient form
\begin{eqnarray}
    \rho^{X}_{Alice} = 
    \biggl(
    \frac{P_{succ}}{2}\ket{0_{x}}\bra{0_{x}} + \frac{1 - P_{succ}}{2}\ket{1_{x}}\bra{1_{x}}
    \biggl) 
    \otimes \ket{0_{\Delta}}\bra{0_{\Delta}} \otimes \ket{r_{0}}\bra{r_{0}}
    + \\ +
    \biggl(
    \frac{1 - P_{succ}}{2}\ket{0_{x}}\bra{0_{x}} + \frac{P_{succ}}{2}\ket{1_{x}}\bra{1_{x}}
    \biggl) 
    \otimes \ket{1_{\Delta}}\bra{1_{\Delta}} \otimes \ket{r_{1}}\bra{r_{1}} + \text{off-diag terms}
\end{eqnarray}
After doing a measurement of ancillary state off-diagonal terms of a compound density matrix vanish. Conditioning on the outcome of the ancilla measurement, the resulting states have the form
\begin{equation}
    \label{reweightedensemble0}
    \text{Measured } \ket{r_{0}}: \rho^{'}_{Alice} = P_{succ}\ket{0_{x}}\bra{0_{x}} + (1 - P_{succ})\ket{1_{x}}\bra{1_{x}}
\end{equation}
\begin{equation}
    \label{reweightedensemble1}
    \text{Measured } \ket{r_{1}}: \rho^{'}_{Alice} = (1 - P_{succ})\ket{0_{x}}\bra{0_{x}} + P_{succ}\ket{1_{x}}\bra{1_{x}}
\end{equation}
The same logic applies to all bases of the BB84 protocol. This leads to reweighting of Alice ensemble states, which is used for adaptive eavesdropping.

%
%

\section{Hong-Ou-Mandel interference}

Legitimate sides can use a Hong-Ou-Mandel interference to keep track of information that leaks to Eve through the photon distinguishability side channel. The Hong-Ou-Mandel interference is a two-photon interference that prohibits two indistinguishable photons exit different ends of a balanced beam splitter. If there is any photon distinguishability in two photons, incident on a beam splitter, there will be a non-zero probability to measure photon counts in both exits of a beam splitter in photodetectors placed on each exit. These coincidence counts alert the physical distinguishability of photons for any physical difference and can be a sign that photons contain additional information available to the eavesdropper.

Let us consider a two-photon state with incident on a balansed beam splitter. The initial state of two photons is 
\begin{equation}
     \ket{\Psi_{in}}_{ab} = a^{\dagger}_{p1}b^{\dagger}_{p2}\ket{0_{z}}_{ab}  = \ket{1; p_{1}}_{a}\ket{1; p_{2}}_{b},
\end{equation}
where $a^{\dagger}$ and $b^{\dagger}$ are creation operators for modes $a$ and $b$, which correspond to income sides of a beam splitter and $p_{1,2}$ are arbitrary discrete degrees of freedom of two photons. The unitary transform of the beam splitter is 
\begin{equation}
    U_{BS}a^{\dagger} = \frac{1}{\sqrt{2}}a^{\dagger} + \frac{1}{\sqrt{2}}b^{\dagger}
\end{equation}
\begin{equation}
    U_{BS}b^{\dagger} = \frac{1}{\sqrt{2}}a^{\dagger} - \frac{1}{\sqrt{2}}b^{\dagger}
\end{equation}
 Applying this operator to the initial two-photon state, one obtains the output state of the form
 \begin{equation}
    \label{twophotonstate}
     \ket{\Psi_{out}}_{ab} = \frac{1}{2}(
     a^{\dagger}_{p1} a^{\dagger}_{p2} + 
     a^{\dagger}_{p2}b^{\dagger}_{p1} - 
     a^{\dagger}_{p1}b^{\dagger}_{p2} - 
     b^{\dagger}_{p1}b^{\dagger}_{p2}
     ) \ket{0_{z}}_{ab}
 \end{equation}
When all degrees of freedom are the same for a pair of incident photons, then this state reduces to
\begin{equation}
     \ket{\Psi_{out}}_{ab} = \frac{1}{2}(
     a^{\dagger}_{p} a^{\dagger}_{p} - 
     b^{\dagger}_{p}b^{\dagger}_{p}
     ) \ket{0_{z}}_{ab}
     =
    \frac{1}{\sqrt{2}}(\ket{2;p})_{a} - \ket{2;p}_{b})
\end{equation}
This state allows two photons to exit one or another exit of the beam splitter in pairs, but not separately. If one places photodetectors on each end of the beam splitter, there will be no coincidence counts (counts of both photodetectors at the same time bin). Contrary, if there is a mismatch in any degrees of freedom of two incident photons, a state (\ref{twophotonstate}) will be generated after the beam splitter, and there will be coincidence counts. These coincidence events will affect the visibility of the HOM interference visibility, defined as 
 \begin{equation}
     V(\rho_{1}, \rho_{2}) = Tr[\rho_{1} \rho_{2}] = \frac{N_{max} - N_{min}}{N_{max}},
 \end{equation}
where $N_{min}$ and $N_{max}$ are minimum and maximum value of coincidence counts, available with varying a particular degree of freedom of two incident photons. If Alice varies her signal degree of freedom (e.g., the polarization of photons) and tests the photons of her source she can test her photons on additional distinguishability with the HOM visibility. If there is no photon distinguishability side channel, the visibility will be equal to unity. If this is not the case and the visibility is less than unity, then there is additional distinguishability among photons, which leaks information to Eve. 

\section{Secret key rates}

In this section, we provide formulae for secret key rates in three eavesdropping strategies demonstrated in the main text.

The first strategy we described is minimal error measurement of the side channel state and a phase covariant cloning of the signal photon state. If Eve makes a minimum error measurement of the side channel, she obtains binary information about the quantum state of the signal photon. This information means that two states inside a basis have different probabilities, and Eve now needs to distinguish between states from a quantum ensemble
\begin{equation}
    \biggl\{ p: \rho_{0}, \quad 1-p: \rho_{1} \biggl\}
\end{equation}
where $p$ and $1-p$ represent a classical knowledge obtained through the side channel. Here Eve uses her classical information twofold: she obtains information from a classical channel and also uses this information in quantum ensemble discrimination. Her total information about secret bits is 
\begin{equation}
    I_{AE} = I_{classical}(p) + I_{quantum}(p,Q) = 1 - h_{2}(p) + 
    h_{2}\biggl(
    \frac{1}{2}(1 - \sqrt{1 - 4p(1-p)(1 - (1 - 2Q)^{2})})
    \biggl)
\end{equation}
where $Q$ is a bit error on the Bob side, and we use the connection between $Q$ and Eve's state distinguishability \cite{Fuchs1997, Brus1998}.

The second strategy we described is minimal error measurement of the side channel with soft filtering and two-state cloning of the signal photon state. Here again Eve has a classical binary information from side channel, but now she uses this information to adjust attack strategy on signal photon. Measurement of side channel leads to reweighting of the ensemble probabilities (see Appendix \ref{AppendixA}): if Eve measures $\ket{0_{\Delta}}$ on the side channel, she obtains ensemble of the form (\ref{reweightedensemble0}), and if she measures $\ket{1_{\Delta}}$, she obtains ensemble of the form (\ref{reweightedensemble1}). Then Eve makes an eavesdropping sequence ``soft-filtering $\longrightarrow$ cloning $\longrightarrow$ backward soft filtering'', and obtains a quantum state, correlated with a secret bit. At the end of communication Eve obtains basis information for every bit position and makes a collective measurement on the ensemble of states in her quantum memory. For an X basis position, Eve discriminates states from an ensemble
\begin{equation}
    \label{evefinalensembleX}
    \mathcal{E} = 
    \biggl\{ 
    P^{0_{\Delta}}_{0}: \rho_{Eve}^{0,0_{\Delta} },
    P^{1_{\Delta}}_{0}: \rho_{Eve}^{0,1_{\Delta} },
    P^{0_{\Delta}}_{1}: \rho_{Eve}^{1,0_{\Delta} },
    P^{1_{\Delta}}_{1}: \rho_{Eve}^{1,1_{\Delta} },
    \biggl\},
\end{equation}
where $P_{0}^{0_{\Delta}}$ is a probability to receive a state $\rho_{Eve}^{0,0_{\Delta} }$. This state is a result of measuring $\ket{0_{\Delta}}$ in side channel, tuning soft filtering on ensemble (\ref{bb84ensemble_q}), postselecting successful filtering results, applying two-state cloning and back soft filtering with postselection of successful results. The same logic applies to other state in the ensemble (\ref{evefinalensembleX}).  
To see, that this ensemble correctly accounts for side channel measurement, we here discuss two extreme cases. If side channel states are orthogonal ($\bra{0_{\Delta}}\ket{1_{\Delta}} = 0$), then this ensemble transforms to
\begin{equation}
    \label{evefinalensembleX}
    \mathcal{E} = 
    \biggl\{ 
    \frac{1}{2}: \ket{0_{x}}\bra{0_{x}},
    0:           \rho_{Eve}^{ 0,1_{\Delta} },
    0:           \rho_{Eve}^{ 1,0_{\Delta} },
    \frac{1}{2}: \ket{1_{x}}\bra{1_{x}},
    \biggl\},
\end{equation}
where states $\rho_{Eve}^{ 0,1_{\Delta} }$ and $\rho_{Eve}^{ 1,0_{\Delta} }$ are some non-orthogonal states (which are not important for calculation since their probability to reach Eve is zero). Here Eve has an ensemble of two equiprobable and orthogonal pure states, which give one bit of information, hence Eve can reliably distinguish a quantum state, sent by Alice in a quantum channel.  Contrary, if side channel states are coincide, then the final ensemble is 
\begin{equation}
    \label{evefinalensembleX}
    \mathcal{E} = 
    \biggl\{ 
    \frac{1}{4}: \rho_{Eve}^{0, 0_{\Delta}},
    \frac{1}{4}: \rho_{Eve}^{0, 1_{\Delta}},
    \frac{1}{4}: \rho_{Eve}^{1, 0_{\Delta}},
    \frac{1}{4}: \rho_{Eve}^{1, 1_{\Delta}},
    \biggl\},
\end{equation}
where $\rho_{Eve}^{0,1}$ are resulting states in Eve quantum memory after the whole attack sequence. The lack of information from side channel (due to complete indistinguishability of side channel states) leads to no effect from soft filtering ((\ref{Fsucc}) is a unity operator) and the only action Eve does on the signal photon state is a two-state cloning, which now is chosen at random. In general case, the secret key rate has a form
\begin{equation}
    R = (1 - P_{attack})\cdot 1 + P_{attack}(1 - h_{2}(Q) - \chi(\mathcal{E}))
\end{equation}
where $Q$ is a bit error on the Bob side. We here introduced a variable $P_{attack}$ to have a controlled parameter for error on Bob side. Here $\chi$ is a Holevo bound value. The Holevo value is a maximal number of bits per state one can extract from the ensemble of quantum states with the best collective measurement of infinite number of states, defined as follows:
\begin{equation}
    \chi(\mathcal{E}) = S(\sum_{j}p_{j}\rho_{j}) - \sum_{j}p_{j}S(\rho_{j}),
\end{equation}
where $S(\rho)$ is the von Neumann entropy
\begin{equation}
    S(\rho) = -Tr[\rho\log_{2}(\rho)].
\end{equation}

The last strategy we described here is a USD measurement of the side channel state with phase-covariant cloning on the signal photon state if the USD measurement failed. Here the secret key \cite{Csiszar1978} rate is 
\begin{equation}
    R = (1 - P_{USD})(1 - h_{2}(Q) - \chi(\mathcal{E}))
\end{equation}
where $P_{USD}$ is a probability of the USD measurement success, $Q$ is a bit error on the Bob side and $\mathcal{E} = \{\frac{1}{2}: \rho_{0,X}, \frac{1}{2}: \rho_{1,X}\}$ is the standard BB84 basis X after phase-covariant cloning eavesdropping.

\bibliography{ref}

\begin{thebibliography}{27}%
\makeatletter
\providecommand \@ifxundefined [1]{%
 \@ifx{#1\undefined}
}%
\providecommand \@ifnum [1]{%
 \ifnum #1\expandafter \@firstoftwo
 \else \expandafter \@secondoftwo
 \fi
}%
\providecommand \@ifx [1]{%
 \ifx #1\expandafter \@firstoftwo
 \else \expandafter \@secondoftwo
 \fi
}%
\providecommand \natexlab [1]{#1}%
\providecommand \enquote  [1]{``#1''}%
\providecommand \bibnamefont  [1]{#1}%
\providecommand \bibfnamefont [1]{#1}%
\providecommand \citenamefont [1]{#1}%
\providecommand \href@noop [0]{\@secondoftwo}%
\providecommand \href [0]{\begingroup \@sanitize@url \@href}%
\providecommand \@href[1]{\@@startlink{#1}\@@href}%
\providecommand \@@href[1]{\endgroup#1\@@endlink}%
\providecommand \@sanitize@url [0]{\catcode `\\12\catcode `\$12\catcode
  `\&12\catcode `\#12\catcode `\^12\catcode `\_12\catcode `\%12\relax}%
\providecommand \@@startlink[1]{}%
\providecommand \@@endlink[0]{}%
\providecommand \url  [0]{\begingroup\@sanitize@url \@url }%
\providecommand \@url [1]{\endgroup\@href {#1}{\urlprefix }}%
\providecommand \urlprefix  [0]{URL }%
\providecommand \Eprint [0]{\href }%
\providecommand \doibase [0]{https://doi.org/}%
\providecommand \selectlanguage [0]{\@gobble}%
\providecommand \bibinfo  [0]{\@secondoftwo}%
\providecommand \bibfield  [0]{\@secondoftwo}%
\providecommand \translation [1]{[#1]}%
\providecommand \BibitemOpen [0]{}%
\providecommand \bibitemStop [0]{}%
\providecommand \bibitemNoStop [0]{.\EOS\space}%
\providecommand \EOS [0]{\spacefactor3000\relax}%
\providecommand \BibitemShut  [1]{\csname bibitem#1\endcsname}%
\let\auto@bib@innerbib\@empty
\bibitem [{\citenamefont {Gisin}\ \emph {et~al.}(2002)\citenamefont {Gisin},
  \citenamefont {Ribordy}, \citenamefont {Tittel},\ and\ \citenamefont
  {Zbinden}}]{gisin2002}%
  \BibitemOpen
  \bibfield  {author} {\bibinfo {author} {\bibfnamefont {N.}~\bibnamefont
  {Gisin}}, \bibinfo {author} {\bibfnamefont {G.}~\bibnamefont {Ribordy}},
  \bibinfo {author} {\bibfnamefont {W.}~\bibnamefont {Tittel}},\ and\ \bibinfo
  {author} {\bibfnamefont {H.}~\bibnamefont {Zbinden}},\ }\bibfield  {title}
  {\bibinfo {title} {Quantum cryptography},\ }\href
  {https://doi.org/10.1103/RevModPhys.74.145} {\bibfield  {journal} {\bibinfo
  {journal} {Rev. Mod. Phys.}\ }\textbf {\bibinfo {volume} {74}},\ \bibinfo
  {pages} {145} (\bibinfo {year} {2002})}\BibitemShut {NoStop}%
\bibitem [{\citenamefont {Pirandola}\ \emph {et~al.}(2020)\citenamefont
  {Pirandola}, \citenamefont {Andersen}, \citenamefont {Banchi}, \citenamefont
  {Berta}, \citenamefont {Bunandar}, \citenamefont {Colbeck}, \citenamefont
  {Englund}, \citenamefont {Gehring}, \citenamefont {Lupo}, \citenamefont
  {Ottaviani}, \citenamefont {Pereira}, \citenamefont {Razavi}, \citenamefont
  {Shaari}, \citenamefont {Tomamichel}, \citenamefont {Usenko}, \citenamefont
  {Vallone}, \citenamefont {Villoresi},\ and\ \citenamefont
  {Wallden}}]{Pirandola2020}%
  \BibitemOpen
  \bibfield  {author} {\bibinfo {author} {\bibfnamefont {S.}~\bibnamefont
  {Pirandola}}, \bibinfo {author} {\bibfnamefont {U.~L.}\ \bibnamefont
  {Andersen}}, \bibinfo {author} {\bibfnamefont {L.}~\bibnamefont {Banchi}},
  \bibinfo {author} {\bibfnamefont {M.}~\bibnamefont {Berta}}, \bibinfo
  {author} {\bibfnamefont {D.}~\bibnamefont {Bunandar}}, \bibinfo {author}
  {\bibfnamefont {R.}~\bibnamefont {Colbeck}}, \bibinfo {author} {\bibfnamefont
  {D.}~\bibnamefont {Englund}}, \bibinfo {author} {\bibfnamefont
  {T.}~\bibnamefont {Gehring}}, \bibinfo {author} {\bibfnamefont
  {C.}~\bibnamefont {Lupo}}, \bibinfo {author} {\bibfnamefont {C.}~\bibnamefont
  {Ottaviani}}, \bibinfo {author} {\bibfnamefont {J.~L.}\ \bibnamefont
  {Pereira}}, \bibinfo {author} {\bibfnamefont {M.}~\bibnamefont {Razavi}},
  \bibinfo {author} {\bibfnamefont {J.~S.}\ \bibnamefont {Shaari}}, \bibinfo
  {author} {\bibfnamefont {M.}~\bibnamefont {Tomamichel}}, \bibinfo {author}
  {\bibfnamefont {V.~C.}\ \bibnamefont {Usenko}}, \bibinfo {author}
  {\bibfnamefont {G.}~\bibnamefont {Vallone}}, \bibinfo {author} {\bibfnamefont
  {P.}~\bibnamefont {Villoresi}},\ and\ \bibinfo {author} {\bibfnamefont
  {P.}~\bibnamefont {Wallden}},\ }\bibfield  {title} {\bibinfo {title}
  {Advances in quantum cryptography},\ }\href
  {https://doi.org/10.1364/AOP.361502} {\bibfield  {journal} {\bibinfo
  {journal} {Adv. Opt. Photon.}\ }\textbf {\bibinfo {volume} {12}},\ \bibinfo
  {pages} {1012} (\bibinfo {year} {2020})}\BibitemShut {NoStop}%
\bibitem [{\citenamefont {Wootters}\ and\ \citenamefont
  {Zurek}(1982)}]{Wooters1982}%
  \BibitemOpen
  \bibfield  {author} {\bibinfo {author} {\bibfnamefont {W.~K.}\ \bibnamefont
  {Wootters}}\ and\ \bibinfo {author} {\bibfnamefont {W.~H.}\ \bibnamefont
  {Zurek}},\ }\bibfield  {title} {\bibinfo {title} {A single quantum cannot be
  cloned},\ }\href {https://doi.org/10.1038/299802a0} {\bibfield  {journal}
  {\bibinfo  {journal} {Nature}\ }\textbf {\bibinfo {volume} {299}},\ \bibinfo
  {pages} {802–803} (\bibinfo {year} {1982})}\BibitemShut {NoStop}%
\bibitem [{\citenamefont {Dieks}(1982)}]{Dieks1982}%
  \BibitemOpen
  \bibfield  {author} {\bibinfo {author} {\bibfnamefont {D.}~\bibnamefont
  {Dieks}},\ }\bibfield  {title} {\bibinfo {title} {Communication by {EPR}
  devices},\ }\href {https://doi.org/10.1016/0375-9601(82)90084-6} {\bibfield
  {journal} {\bibinfo  {journal} {Physics Letters A}\ }\textbf {\bibinfo
  {volume} {92}},\ \bibinfo {pages} {271} (\bibinfo {year} {1982})}\BibitemShut
  {NoStop}%
\bibitem [{\citenamefont {Lo}\ and\ \citenamefont {Chau}(1999)}]{LoChau1999}%
  \BibitemOpen
  \bibfield  {author} {\bibinfo {author} {\bibfnamefont {H.-K.}\ \bibnamefont
  {Lo}}\ and\ \bibinfo {author} {\bibfnamefont {H.~F.}\ \bibnamefont {Chau}},\
  }\bibfield  {title} {\bibinfo {title} {Unconditional security of quantum key
  distribution over arbitrarily long distances},\ }\href
  {https://doi.org/10.1126/science.283.5410.2050} {\bibfield  {journal}
  {\bibinfo  {journal} {Science}\ }\textbf {\bibinfo {volume} {283}},\ \bibinfo
  {pages} {2050} (\bibinfo {year} {1999})}\BibitemShut {NoStop}%
\bibitem [{\citenamefont {Shor}\ and\ \citenamefont
  {Preskill}(2000)}]{Shor2000}%
  \BibitemOpen
  \bibfield  {author} {\bibinfo {author} {\bibfnamefont {P.~W.}\ \bibnamefont
  {Shor}}\ and\ \bibinfo {author} {\bibfnamefont {J.}~\bibnamefont
  {Preskill}},\ }\bibfield  {title} {\bibinfo {title} {Simple proof of security
  of the {BB}84 quantum key distribution protocol},\ }\href
  {https://doi.org/10.1103/PhysRevLett.85.441} {\bibfield  {journal} {\bibinfo
  {journal} {Phys. Rev. Lett.}\ }\textbf {\bibinfo {volume} {85}},\ \bibinfo
  {pages} {441} (\bibinfo {year} {2000})}\BibitemShut {NoStop}%
\bibitem [{\citenamefont {Scarani}\ and\ \citenamefont
  {Kurtsiefer}(2014)}]{Scarani2014}%
  \BibitemOpen
  \bibfield  {author} {\bibinfo {author} {\bibfnamefont {V.}~\bibnamefont
  {Scarani}}\ and\ \bibinfo {author} {\bibfnamefont {C.}~\bibnamefont
  {Kurtsiefer}},\ }\bibfield  {title} {\bibinfo {title} {The black paper of
  quantum cryptography: real implementation problems},\ }\href
  {https://doi.org/10.1016/j.tcs.2014.09.015} {\bibfield  {journal} {\bibinfo
  {journal} {Rev. Mod. Phys.}\ }\textbf {\bibinfo {volume} {560}},\ \bibinfo
  {pages} {27} (\bibinfo {year} {2014})}\BibitemShut {NoStop}%
\bibitem [{\citenamefont {Jain}\ \emph {et~al.}(2016)\citenamefont {Jain},
  \citenamefont {Stiller}, \citenamefont {Khan}, \citenamefont {Elser},
  \citenamefont {Marquardt},\ and\ \citenamefont {Leuchs}}]{Jain2016}%
  \BibitemOpen
  \bibfield  {author} {\bibinfo {author} {\bibfnamefont {N.}~\bibnamefont
  {Jain}}, \bibinfo {author} {\bibfnamefont {B.}~\bibnamefont {Stiller}},
  \bibinfo {author} {\bibfnamefont {I.}~\bibnamefont {Khan}}, \bibinfo {author}
  {\bibfnamefont {D.}~\bibnamefont {Elser}}, \bibinfo {author} {\bibfnamefont
  {C.}~\bibnamefont {Marquardt}},\ and\ \bibinfo {author} {\bibfnamefont
  {G.}~\bibnamefont {Leuchs}},\ }\bibfield  {title} {\bibinfo {title} {Attacks
  on practical quantum key distribution systems (and how to prevent them)},\
  }\href {https://doi.org/10.1080/00107514.2016.1148333} {\bibfield  {journal}
  {\bibinfo  {journal} {Contemporary Physics}\ }\textbf {\bibinfo {volume}
  {57}},\ \bibinfo {pages} {366} (\bibinfo {year} {2016})}\BibitemShut
  {NoStop}%
\bibitem [{\citenamefont {Mar{\o}y}\ \emph {et~al.}(2017)\citenamefont
  {Mar{\o}y}, \citenamefont {Makarov},\ and\ \citenamefont
  {Skaar}}]{Makarov2017}%
  \BibitemOpen
  \bibfield  {author} {\bibinfo {author} {\bibfnamefont {{\O}.}~\bibnamefont
  {Mar{\o}y}}, \bibinfo {author} {\bibfnamefont {V.}~\bibnamefont {Makarov}},\
  and\ \bibinfo {author} {\bibfnamefont {J.}~\bibnamefont {Skaar}},\ }\bibfield
   {title} {\bibinfo {title} {Secure detection in quantum key distribution by
  real-time calibration of receiver},\ }\href
  {https://doi.org/10.1088/2058-9565/aa83c9} {\bibfield  {journal} {\bibinfo
  {journal} {Quantum Science and Technology}\ }\textbf {\bibinfo {volume}
  {2}},\ \bibinfo {pages} {044013} (\bibinfo {year} {2017})}\BibitemShut
  {NoStop}%
\bibitem [{\citenamefont {Huang}\ \emph {et~al.}(2018)\citenamefont {Huang},
  \citenamefont {Sun}, \citenamefont {Liu},\ and\ \citenamefont
  {Makarov}}]{Huang2018}%
  \BibitemOpen
  \bibfield  {author} {\bibinfo {author} {\bibfnamefont {A.}~\bibnamefont
  {Huang}}, \bibinfo {author} {\bibfnamefont {S.-H.}\ \bibnamefont {Sun}},
  \bibinfo {author} {\bibfnamefont {Z.}~\bibnamefont {Liu}},\ and\ \bibinfo
  {author} {\bibfnamefont {V.}~\bibnamefont {Makarov}},\ }\bibfield  {title}
  {\bibinfo {title} {Quantum key distribution with distinguishable decoy
  states},\ }\href {https://doi.org/10.1103/PhysRevA.98.012330} {\bibfield
  {journal} {\bibinfo  {journal} {Phys. Rev. A}\ }\textbf {\bibinfo {volume}
  {98}},\ \bibinfo {pages} {012330} (\bibinfo {year} {2018})}\BibitemShut
  {NoStop}%
\bibitem [{\citenamefont {Diamanti}\ \emph {et~al.}(2016)\citenamefont
  {Diamanti}, \citenamefont {Lo}, \citenamefont {Qi},\ and\ \citenamefont
  {Yuan}}]{Diamanti2016}%
  \BibitemOpen
  \bibfield  {author} {\bibinfo {author} {\bibfnamefont {E.}~\bibnamefont
  {Diamanti}}, \bibinfo {author} {\bibfnamefont {H.-K.}\ \bibnamefont {Lo}},
  \bibinfo {author} {\bibfnamefont {B.}~\bibnamefont {Qi}},\ and\ \bibinfo
  {author} {\bibfnamefont {Z.}~\bibnamefont {Yuan}},\ }\bibfield  {title}
  {\bibinfo {title} {Practical challenges in quantum key distribution},\ }\href
  {https://doi.org/10.1038/npjqi.2016.25} {\bibfield  {journal} {\bibinfo
  {journal} {npj Quantum Information}\ }\textbf {\bibinfo {volume} {2}}
  (\bibinfo {year} {2016})}\BibitemShut {NoStop}%
\bibitem [{\citenamefont {Xu}\ \emph {et~al.}(2020)\citenamefont {Xu},
  \citenamefont {Ma}, \citenamefont {Zhang}, \citenamefont {Lo},\ and\
  \citenamefont {Pan}}]{Xu2020}%
  \BibitemOpen
  \bibfield  {author} {\bibinfo {author} {\bibfnamefont {F.}~\bibnamefont
  {Xu}}, \bibinfo {author} {\bibfnamefont {X.}~\bibnamefont {Ma}}, \bibinfo
  {author} {\bibfnamefont {Q.}~\bibnamefont {Zhang}}, \bibinfo {author}
  {\bibfnamefont {H.-K.}\ \bibnamefont {Lo}},\ and\ \bibinfo {author}
  {\bibfnamefont {J.-W.}\ \bibnamefont {Pan}},\ }\bibfield  {title} {\bibinfo
  {title} {Secure quantum key distribution with realistic devices},\ }\href
  {https://doi.org/10.1103/RevModPhys.92.025002} {\bibfield  {journal}
  {\bibinfo  {journal} {Rev. Mod. Phys.}\ }\textbf {\bibinfo {volume} {92}},\
  \bibinfo {pages} {025002} (\bibinfo {year} {2020})}\BibitemShut {NoStop}%
\bibitem [{\citenamefont {Lo}\ \emph {et~al.}(2012)\citenamefont {Lo},
  \citenamefont {Curty},\ and\ \citenamefont {Qi}}]{MDI}%
  \BibitemOpen
  \bibfield  {author} {\bibinfo {author} {\bibfnamefont {H.-K.}\ \bibnamefont
  {Lo}}, \bibinfo {author} {\bibfnamefont {M.}~\bibnamefont {Curty}},\ and\
  \bibinfo {author} {\bibfnamefont {B.}~\bibnamefont {Qi}},\ }\bibfield
  {title} {\bibinfo {title} {Measurement-device-independent quantum key
  distribution},\ }\href {https://doi.org/10.1103/PhysRevLett.108.130503}
  {\bibfield  {journal} {\bibinfo  {journal} {Phys. Rev. Lett.}\ }\textbf
  {\bibinfo {volume} {108}},\ \bibinfo {pages} {130503} (\bibinfo {year}
  {2012})}\BibitemShut {NoStop}%
\bibitem [{\citenamefont {Lucamarini}\ \emph {et~al.}(2015)\citenamefont
  {Lucamarini}, \citenamefont {Choi}, \citenamefont {Ward}, \citenamefont
  {Dynes}, \citenamefont {Yuan},\ and\ \citenamefont
  {Shields}}]{Lucamarini2015}%
  \BibitemOpen
  \bibfield  {author} {\bibinfo {author} {\bibfnamefont {M.}~\bibnamefont
  {Lucamarini}}, \bibinfo {author} {\bibfnamefont {I.}~\bibnamefont {Choi}},
  \bibinfo {author} {\bibfnamefont {M.~B.}\ \bibnamefont {Ward}}, \bibinfo
  {author} {\bibfnamefont {J.~F.}\ \bibnamefont {Dynes}}, \bibinfo {author}
  {\bibfnamefont {Z.~L.}\ \bibnamefont {Yuan}},\ and\ \bibinfo {author}
  {\bibfnamefont {A.~J.}\ \bibnamefont {Shields}},\ }\bibfield  {title}
  {\bibinfo {title} {Practical security bounds against the {T}rojan-horse
  attack in quantum key distribution},\ }\href
  {https://doi.org/10.1103/PhysRevX.5.031030} {\bibfield  {journal} {\bibinfo
  {journal} {Phys. Rev. X}\ }\textbf {\bibinfo {volume} {5}},\ \bibinfo {pages}
  {031030} (\bibinfo {year} {2015})}\BibitemShut {NoStop}%
\bibitem [{\citenamefont {Tamaki}\ \emph {et~al.}(2016)\citenamefont {Tamaki},
  \citenamefont {Curty},\ and\ \citenamefont {Lucamarini}}]{Tamaki2016}%
  \BibitemOpen
  \bibfield  {author} {\bibinfo {author} {\bibfnamefont {K.}~\bibnamefont
  {Tamaki}}, \bibinfo {author} {\bibfnamefont {M.}~\bibnamefont {Curty}},\ and\
  \bibinfo {author} {\bibfnamefont {M.}~\bibnamefont {Lucamarini}},\ }\bibfield
   {title} {\bibinfo {title} {Decoy-state quantum key distribution with a leaky
  source},\ }\href {https://doi.org/10.1088/1367-2630/18/6/065008} {\bibfield
  {journal} {\bibinfo  {journal} {New Journal of Physics}\ }\textbf {\bibinfo
  {volume} {18}},\ \bibinfo {pages} {065008} (\bibinfo {year}
  {2016})}\BibitemShut {NoStop}%
\bibitem [{\citenamefont {Nauerth}\ \emph {et~al.}(2009)\citenamefont
  {Nauerth}, \citenamefont {Fürst}, \citenamefont {Schmitt-Manderbach},
  \citenamefont {Weier},\ and\ \citenamefont {Weinfurter}}]{Nauerth2009}%
  \BibitemOpen
  \bibfield  {author} {\bibinfo {author} {\bibfnamefont {S.}~\bibnamefont
  {Nauerth}}, \bibinfo {author} {\bibfnamefont {M.}~\bibnamefont {Fürst}},
  \bibinfo {author} {\bibfnamefont {T.}~\bibnamefont {Schmitt-Manderbach}},
  \bibinfo {author} {\bibfnamefont {H.}~\bibnamefont {Weier}},\ and\ \bibinfo
  {author} {\bibfnamefont {H.}~\bibnamefont {Weinfurter}},\ }\bibfield  {title}
  {\bibinfo {title} {Information leakage via side channels in freespace {BB}84
  quantum cryptography},\ }\href
  {https://doi.org/10.1088/1367-2630/11/6/065001} {\bibfield  {journal}
  {\bibinfo  {journal} {New Journal of Physics}\ }\textbf {\bibinfo {volume}
  {11}},\ \bibinfo {pages} {065001} (\bibinfo {year} {2009})}\BibitemShut
  {NoStop}%
\bibitem [{\citenamefont {Duplinskiy}\ and\ \citenamefont
  {Sych}(2021)}]{Duplinskii2019}%
  \BibitemOpen
  \bibfield  {author} {\bibinfo {author} {\bibfnamefont {A.}~\bibnamefont
  {Duplinskiy}}\ and\ \bibinfo {author} {\bibfnamefont {D.}~\bibnamefont
  {Sych}},\ }\bibfield  {title} {\bibinfo {title} {Bounding light source side
  channels in {QKD} via {H}ong-{O}u-{M}andel interference},\ }\href
  {https://doi.org/10.1103/PhysRevA.104.012601} {\bibfield  {journal} {\bibinfo
   {journal} {Phys. Rev. A}\ }\textbf {\bibinfo {volume} {104}},\ \bibinfo
  {pages} {012601} (\bibinfo {year} {2021})}\BibitemShut {NoStop}%
\bibitem [{\citenamefont {Pereira}\ \emph {et~al.}(2019)\citenamefont
  {Pereira}, \citenamefont {Curty},\ and\ \citenamefont
  {Tamaki}}]{Pereira2019}%
  \BibitemOpen
  \bibfield  {author} {\bibinfo {author} {\bibfnamefont {M.}~\bibnamefont
  {Pereira}}, \bibinfo {author} {\bibfnamefont {M.}~\bibnamefont {Curty}},\
  and\ \bibinfo {author} {\bibfnamefont {K.}~\bibnamefont {Tamaki}},\
  }\bibfield  {title} {\bibinfo {title} {Quantum key distribution with flawed
  and leaky sources},\ }\href {https://doi.org/10.1038/s41534-019-0180-9}
  {\bibfield  {journal} {\bibinfo  {journal} {npj Quantum Information}\
  }\textbf {\bibinfo {volume} {5}} (\bibinfo {year} {2019})}\BibitemShut
  {NoStop}%
\bibitem [{\citenamefont {Bruß}\ \emph {et~al.}(2000)\citenamefont {Bruß},
  \citenamefont {David P.~DiVincenzo}, \citenamefont {Fuchs}, \citenamefont
  {Macchiavello},\ and\ \citenamefont {Smolin}}]{Brus2000}%
  \BibitemOpen
  \bibfield  {author} {\bibinfo {author} {\bibfnamefont {D.}~\bibnamefont
  {Bruß}}, \bibinfo {author} {\bibfnamefont {A.~E.}\ \bibnamefont {David
  P.~DiVincenzo}}, \bibinfo {author} {\bibfnamefont {C.~A.}\ \bibnamefont
  {Fuchs}}, \bibinfo {author} {\bibfnamefont {C.}~\bibnamefont
  {Macchiavello}},\ and\ \bibinfo {author} {\bibfnamefont {J.~A.}\ \bibnamefont
  {Smolin}},\ }\bibfield  {title} {\bibinfo {title} {Phase-covariant quantum
  cloning},\ }\href {https://doi.org/10.1103/PhysRevA.57.2368} {\bibfield
  {journal} {\bibinfo  {journal} {Phys. Rev. A}\ }\textbf {\bibinfo {volume}
  {62}},\ \bibinfo {pages} {012302} (\bibinfo {year} {2000})}\BibitemShut
  {NoStop}%
\bibitem [{\citenamefont {Babukhin}\ and\ \citenamefont
  {Sych}(2020)}]{Babukhin2020}%
  \BibitemOpen
  \bibfield  {author} {\bibinfo {author} {\bibfnamefont {D.}~\bibnamefont
  {Babukhin}}\ and\ \bibinfo {author} {\bibfnamefont {D.}~\bibnamefont
  {Sych}},\ }\bibfield  {title} {\bibinfo {title} {Intercept-resend attack on
  passive side channel of the light source in {BB}84 decoy-state protocol},\
  }\href {https://doi.org/10.1088/1742-6596/1695/1/012119} {\bibfield
  {journal} {\bibinfo  {journal} {Journal of Physics: Conference Series}\
  }\textbf {\bibinfo {volume} {1695}},\ \bibinfo {pages} {012119} (\bibinfo
  {year} {2020})}\BibitemShut {NoStop}%
\bibitem [{\citenamefont {Babukhin}\ and\ \citenamefont
  {Sych}(2021)}]{Babukhin2021}%
  \BibitemOpen
  \bibfield  {author} {\bibinfo {author} {\bibfnamefont {D.}~\bibnamefont
  {Babukhin}}\ and\ \bibinfo {author} {\bibfnamefont {D.}~\bibnamefont
  {Sych}},\ }\bibfield  {title} {\bibinfo {title} {Explicit attacks on passive
  side channels of the light source in the {BB}84 decoy state protocol},\
  }\href {https://doi.org/10.1088/1742-6596/1984/1/012008} {\bibfield
  {journal} {\bibinfo  {journal} {Journal of Physics: Conference Series}\
  }\textbf {\bibinfo {volume} {1984}},\ \bibinfo {pages} {012008} (\bibinfo
  {year} {2021})}\BibitemShut {NoStop}%
\bibitem [{\citenamefont {Sych}\ \emph {et~al.}(2021)\citenamefont {Sych},
  \citenamefont {Duplinskiy},\ and\ \citenamefont {Babukhin}}]{Sych2021}%
  \BibitemOpen
  \bibfield  {author} {\bibinfo {author} {\bibfnamefont {D.}~\bibnamefont
  {Sych}}, \bibinfo {author} {\bibfnamefont {A.}~\bibnamefont {Duplinskiy}},\
  and\ \bibinfo {author} {\bibfnamefont {D.}~\bibnamefont {Babukhin}},\
  }\bibfield  {title} {\bibinfo {title} {Practical security of quantum key
  distribution in the presence of side channels},\ }\href
  {https://doi.org/10.1088/1742-6596/1984/1/012001} {\bibfield  {journal}
  {\bibinfo  {journal} {Journal of Physics: Conference Series}\ }\textbf
  {\bibinfo {volume} {1984}},\ \bibinfo {pages} {012001} (\bibinfo {year}
  {2021})}\BibitemShut {NoStop}%
\bibitem [{\citenamefont {Kronberg}\ \emph {et~al.}(2020)\citenamefont
  {Kronberg}, \citenamefont {Nikolaeva}, \citenamefont {Kurochkin},\ and\
  \citenamefont {Fedorov}}]{kronberg2020quantum}%
  \BibitemOpen
  \bibfield  {author} {\bibinfo {author} {\bibfnamefont {D.}~\bibnamefont
  {Kronberg}}, \bibinfo {author} {\bibfnamefont {A.}~\bibnamefont {Nikolaeva}},
  \bibinfo {author} {\bibfnamefont {Y.~V.}\ \bibnamefont {Kurochkin}},\ and\
  \bibinfo {author} {\bibfnamefont {A.}~\bibnamefont {Fedorov}},\ }\bibfield
  {title} {\bibinfo {title} {Quantum soft filtering for the improved security
  analysis of the coherent one-way quantum-key-distribution protocol},\
  }\href@noop {} {\bibfield  {journal} {\bibinfo  {journal} {Physical Review
  A}\ }\textbf {\bibinfo {volume} {101}},\ \bibinfo {pages} {032334} (\bibinfo
  {year} {2020})}\BibitemShut {NoStop}%
\bibitem [{\citenamefont {Kronberg}(2021)}]{kronberg2021increasing}%
  \BibitemOpen
  \bibfield  {author} {\bibinfo {author} {\bibfnamefont {D.}~\bibnamefont
  {Kronberg}},\ }\bibfield  {title} {\bibinfo {title} {Increasing the
  distinguishability of quantum states with an arbitrary success probability},\
  }\href@noop {} {\bibfield  {journal} {\bibinfo  {journal} {Proceedings of the
  Steklov Institute of Mathematics}\ }\textbf {\bibinfo {volume} {313}},\
  \bibinfo {pages} {113} (\bibinfo {year} {2021})}\BibitemShut {NoStop}%
\bibitem [{\citenamefont {Bruß}\ \emph {et~al.}(1998)\citenamefont {Bruß},
  \citenamefont {DiVincenzo}, \citenamefont {Ekert}, \citenamefont {Fuchs},
  \citenamefont {Macchiavello},\ and\ \citenamefont {Smolin}}]{Brus1998}%
  \BibitemOpen
  \bibfield  {author} {\bibinfo {author} {\bibfnamefont {D.}~\bibnamefont
  {Bruß}}, \bibinfo {author} {\bibfnamefont {D.~P.}\ \bibnamefont
  {DiVincenzo}}, \bibinfo {author} {\bibfnamefont {A.}~\bibnamefont {Ekert}},
  \bibinfo {author} {\bibfnamefont {C.~A.}\ \bibnamefont {Fuchs}}, \bibinfo
  {author} {\bibfnamefont {C.}~\bibnamefont {Macchiavello}},\ and\ \bibinfo
  {author} {\bibfnamefont {J.~A.}\ \bibnamefont {Smolin}},\ }\bibfield  {title}
  {\bibinfo {title} {Optimal universal and state-dependent quantum cloning},\
  }\href {https://doi.org/10.1103/PhysRevA.57.2368} {\bibfield  {journal}
  {\bibinfo  {journal} {Phys. Rev. A}\ }\textbf {\bibinfo {volume} {57}},\
  \bibinfo {pages} {2368} (\bibinfo {year} {1998})}\BibitemShut {NoStop}%
\bibitem [{\citenamefont {Fuchs}\ \emph {et~al.}(1997)\citenamefont {Fuchs},
  \citenamefont {Gisin}, \citenamefont {Griffiths}, \citenamefont {Niu},\ and\
  \citenamefont {Peres}}]{Fuchs1997}%
  \BibitemOpen
  \bibfield  {author} {\bibinfo {author} {\bibfnamefont {C.~A.}\ \bibnamefont
  {Fuchs}}, \bibinfo {author} {\bibfnamefont {N.}~\bibnamefont {Gisin}},
  \bibinfo {author} {\bibfnamefont {R.~B.}\ \bibnamefont {Griffiths}}, \bibinfo
  {author} {\bibfnamefont {C.-S.}\ \bibnamefont {Niu}},\ and\ \bibinfo {author}
  {\bibfnamefont {A.}~\bibnamefont {Peres}},\ }\bibfield  {title} {\bibinfo
  {title} {Optimal eavesdropping in quantum cryptography. i. information bound
  and optimal strategy},\ }\href@noop {} {\bibfield  {journal} {\bibinfo
  {journal} {Physical Review A}\ }\textbf {\bibinfo {volume} {56}},\ \bibinfo
  {pages} {1163} (\bibinfo {year} {1997})}\BibitemShut {NoStop}%
\bibitem [{\citenamefont {Csiszar}\ and\ \citenamefont
  {Korner}(1978)}]{Csiszar1978}%
  \BibitemOpen
  \bibfield  {author} {\bibinfo {author} {\bibfnamefont {I.}~\bibnamefont
  {Csiszar}}\ and\ \bibinfo {author} {\bibfnamefont {J.}~\bibnamefont
  {Korner}},\ }\bibfield  {title} {\bibinfo {title} {Broadcast channels with
  confidential messages},\ }\href {https://doi.org/10.1109/tit.1978.1055892}
  {\bibfield  {journal} {\bibinfo  {journal} {{IEEE} Transactions on
  Information Theory}\ }\textbf {\bibinfo {volume} {24}},\ \bibinfo {pages}
  {339} (\bibinfo {year} {1978})}\BibitemShut {NoStop}%
\end{thebibliography}%


%
\end{document}